\documentclass{article}

\usepackage[utf8]{inputenc}
\usepackage{natbib}
\usepackage{graphicx}
\usepackage{amssymb,bbm}
\usepackage{siunitx}

\title{Wildfires vegetation recovery through satellite remote sensing and 
Functional Data Analysis}
\author{
Serra-Burriel, Feliu \\
 \texttt{feliu.serra@bsc.es}
 \and
Delicado, Pedro \\
 \texttt{pedro.delicado@upc.edu}
 \and
 Cucchietti, Fernando \\
 \texttt{fernando.cucchietti@bsc.es}
}
\date{February 2021}
\renewcommand{\baselinestretch}{1.5}

\begin{document}

\maketitle

\abstract{In recent years wildfires have caused havoc across the world, especially aggravated in certain regions, due to climate change.
Remote sensing has become a powerful tool for monitoring fires, as well as for measuring their effects on vegetation over the following years.
We aim to explain the dynamics of wildfires' effects on a vegetation index (previously estimated by causal inference through synthetic controls) from pre-wildfire available information (mainly proceeding from satellites). 
For this purpose, we use regression models from Functional Data Analysis, where wildfire effects are considered functional responses, depending on elapsed time after each wildfire, while pre-wildfire information acts as scalar covariates. 
Our main findings show that vegetation recovery after wildfires is a slow process, affected by many pre-wildfire conditions, among which the richness and diversity of vegetation is one of the best predictors for the recovery. \\
% Keywords
%\keyword{
}

%%%%%%%%%%%%%%%%%%%%%%%%%%%%%%%%%%%%%%%%%%
\section{Introduction}
% Context and why is this relevant? (Motivation)
Wildfires are becoming a major concern for societies around the globe, and research shows that changes in climate are going to alter the amount and size of wildfires in specific regions \citep{Spracklen2009, Bryant2014, Westerling2011, Westerling2016}. The effects are diverse depending on many factors, like weather conditions, vegetation affected, land cover, land management before and after the incident, the geographical region affected, or human vegetation management and risk mitigation. Wildfires occur by a combination of conditions created either by human intervention (e.g. power lines failures \citep{MITCHELL2013726}) or by unpredictable events (such as lightnings \citep{keeley1982distribution, AMATULLI2007321}, and thus are much harder to anticipate). As natural environments become more vulnerable to this kind of events, cities and inhabitable places need to be made more resilient, as they are likely to become more frequent due to changes in climate \citep{Westerling2011}.
The result of these events increasing in size and frequency is hard to capture, as the amount of ecosystems and populations affected by these is very large. 

Remote sensing can be defined as \textit{``the science of observation from a distance"} \citep{barrett1999introduction}, including many types of sensors. In this study, we are particularly interested in satellite images. These have become an invaluable and increasingly popular research field of study in the last few decades. 
Observation of the Earth from a distance has enormous potential.
It allows monitoring and capturing changes in environments around the world, enabling their detection, quantification and possible prevention, which makes the modification of human environments more sustainable.
Historically, natural disasters have played an important role in shaping societies, as these pose a significant threat in some regions on Earth. In order to create resilient and sustainable communities, remote sensing tools can help adapt to these events \citep{Scheffer2009, Verbesselt2016, Liu2019}, and help build environmental policies to protect Earth as we know it \citep{DeLeeuw2010}. In this work, we are focusing on how wildfires affect vegetation and how environments recover from these catastrophic events. Remote sensing plays a critical role for assessing the impact of wildfires and learning to coexist with these events \citep{Moritz2014}.
We use Functional Data Analysis (FDA, \cite{RamSil:2005}) to analyze wildfire dynamics from remote sensing data.
This work is part of the growing literature on FDA for remote sensing data 
(see, e.g., \cite{acar2018functional}, \cite{militino2019filling} or \cite{sugianto2019functional} among others).

Information from Remote Sensing provides a very important temporal component that allows studying and quantifying the dynamical evolution of the effects of wildfires and recoveries over time. Precisely, \citep{serra2020estimating} use various sources of remote sensing data, combined with synthetic controls for assessing the vegetation impacts of wildfires over time. 
In this study, we analyze these recoveries processes as functional data. 
Each observation measures over several years how the vegetation evolves in a specific region that suffered from a large wildfire,
and it represents the decrease or loss of vegetation (that will be defined in the next sections) from each wildfire, as a function of time $t$, starting at the time of the wildfire up until $7$ years after the wildfire, showing the recovery of vegetation from these events.

% Present methods and main results
Hence, the aim of this study is to explain the effects of wildfires on vegetation from remote sensing (satellite) images through FDA, as an alternative approach to classical regression methodologies used to study the effects of wildfires. Classical models usually summarize the whole recovery by comparing few periods of time, pre- and post-wildfire \citep{Engel2011}. 
We take advantage of remote sensing technologies and modern statistical tools to answer questions like the following: 
\textit{i)} What are the effects of wildfires on different kinds of environments? 
\textit{ii)} Do the wildfire effects evolution depend on the vegetation of the burned area? 
\textit{iii)} Can we explain recoveries of vegetation from wildfires using pre-wildfire observable covariates?

This study is focused on medium to large wildfires ($\geq 1000$ acres, or $404$ hectares) in California throughout a time-span of two decades (1996-2016). We explain the recoveries of vegetation from wildfires using pre-wildfire vegetation conditions and other characteristics of the affected lands using FDA. 
One of the main advantages from this methodology is that we can use the whole recovery process as a function of time.
%, as opposed to other classical regression methods that summarize the whole recovery by comparing two periods of time, pre- and post-wildfire.

Previous studies use differences between pre- and post-wildfire occurrence, showing relative difference between values over fixed time periods, or comparisons of few wildfires (e.g. a dozen wildfires \citep{Bright2019}, 3 or 5 years after the event\citep{Casady2010, Steiner2020}). This results in raster maps of differences between few time periods, gaining insights on the exact locations where vegetation has decreased. However, this approach lacks the temporal nature of the problem, as vegetation changes over time in a continuous manner. 

In order to estimate the dynamical causal effects of wildfires, causal inference through synthetic controls was used in \citep{serra2020estimating}. 
This methodology comes from the combination of Econometrics and Political Science, and it consists on the estimation of a hypothetical scenario (a counterfactual) with the absence of a wildfire (the intervention). Thus, in the present case, health vegetation indices were estimated in places where there were wildfires, as if the wildfires had not happened, using a Generalized Synthetic Control (GSC) methodology \citep{Xu2017}.
Then, the wildfire effect was estimated as the difference between the observed indices and the estimated counterfactuals.
Usually the size of the wildfire effect decreases over time so we also refer as {\em wildfire recovery} to the wildfire effect as a function of time.

We use seasonality adjustment techniques to extract the trend of the wildfire effects estimated in the previous study.
Then proceed to regress these effects, measured over time, using Functional Regression Models. 
More precisely, we regress functional responses on scalar covariates. This results in estimated coefficients changing over time that provide insights into different questions, as the ones stated above.   

This paper is structured as follows. 
First, we introduce the used data and their pre-processing, as well as the algorithms used to obtain the outcomes to be predicted. 
Next, we explain the methodology that will be used in this study. 
Then, we show the attained results and summarize the key findings derived from this study. 
Last, we discuss the potential impact of these results and conclude with final notes.

\section{Data Gathering}
The study area of this paper is California over the time span 1996-2016. There are three main data sources used for this study. First, perimeters from large wildfires ($\geq 404$ hectares) %that occurred between 1984-2016
were obtained from the Monitoring Trends in Burn Severity (MTBS) program \citep{Eidenshink2007} conducted by the United States Geological Services (USGS). 
Second, the Normalized Difference Vegetation Index (NDVI) Surface-Reflectances coming from several Landsat satellites was derived and aggregated using Google Earth Engine platform (GEE) over the areas of interest, as well as meteorological conditions over the areas of interest, that were obtained from GridMET \citep{Abatzoglou2013} during the observed time span. Third, we use the results from a previous analysis in \citep{serra2020estimating}, where the effects of wildfires were estimated using the above two mentioned data  sources. 
Details on these data sources are expanded below.

\subsection{Wildfires Data}
\begin{figure}
    \centering
\hspace*{-1.5cm}
    \includegraphics[width=.85\textwidth]{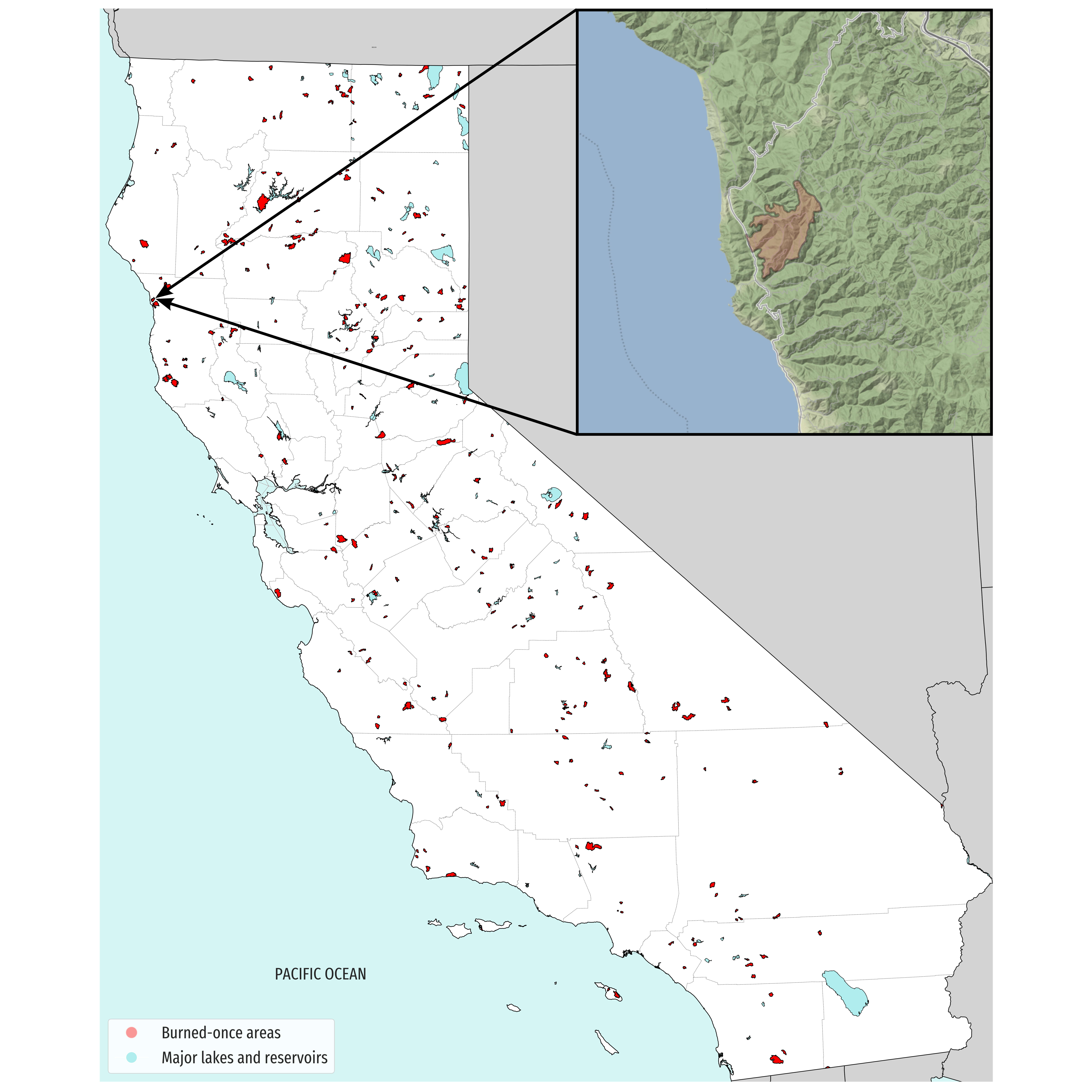}
    \caption{Map of California and the perimeters selected for this study. The upper right corner shows the perimeter of the MEU LIGHTNING COMPLEX (MIDDLE) wildfire in June 2008.}
    \label{fig:California_Map_FDA}
\end{figure}

Perimeters from large wildfires ($\geq 404$ hectares) that occurred over the considered time span were obtained from  MTBS \citep{Eidenshink2007} program, as it provides a consistent source of wildfire perimeters for this period. 
Additionally, 
only perimeters of wildfires that didn't overlap each other over the time period studied have been considered, because the synthetic control methodology used in \citep{serra2020estimating} is not able to deal with units that  experiment more than one intervention (multiple wildfires, in this case).
After pruning the wildfires that either occurred too early and thus don't have enough pre-wildfire periods (at least 5 years) to estimate the counterfactual vegetation, and the wildfires that do not have enough follow-up years after the wildfire (at least 7 years), we end up with 243 wildfires. Figure \ref{fig:California_Map_FDA} shows the perimeters of the burned areas. 
As an example, the upper right corner of Figure \ref{fig:California_Map_FDA} 
shows the perimeter from a 2008 wildfire in the Mendocino County, officially named \textit{MEU LIGHTNING COMPLEX (MIDDLE)}. 
This fire burned 2087 acres, and the predominant land cover was evergreen forest.
We have chosen this wildfire as an example because it corresponds to 
the modal median for 2008 (the deepest function in 2008 according to the modal depth \citep{cuevas2007robust}) and 2008 was the year with the largest amount of wildfires.

Moreover, several spatial covariates were obtained from MTBS:  latitudinal and longitudinal centroid of the polygons, the year that the fire occurred, the month when it started, and the acres or size (in acres) of the burned areas. Lastly, another covariate indicating the average elevation of the burned areas was obtained from the National Elevation Dataset (NED) from the USGS. Table \ref{tab:list_of_covariates} shows a summary of the used covariates in this study.

\renewcommand{\baselinestretch}{.8} 
\renewcommand{\arraystretch}{1.5}
\begin{table}%[H]
     \caption{Summary of the variables used in this study as covariates for the function-on-scalar regressions.}
    \label{tab:list_of_covariates}
  \centering
    \begin{tabular}{r p{.45\textwidth} c}
    \textbf{Variable} & {\bf Description} & {\bf Source} \\
    \hline
  Latitude & Average of the South-North latitude coordinates for the pixels in the area of interest. & MTBS\\
  Longitude & Average of the West-East longitude coordinates for the pixels in the area of interest. & MTBS\\ 
  Avg Elevation & Average of the elevation over the sea level for the pixels in the area of interest.& NED \\ 
  Year & Year the wildfire occurred. & MTBS\\ 
  Start Month & Month the wildfire started. & MTBS\\ 
  log(Acres) & Logarithm of the surface (in acres) of the burned area. & MTBS\\ 
  Landcover & Predominant type of vegetation over the area of interest. Four categories: Shrubland/scrubland, evergreen forest, grasslands herbaceous and others. & GlobCover\\ 
  Landcover Entropy & Shannon's Entropy of the distribution of Landcover among the pixels in the area of interest. Larger values indicate more variety of vegetation types. & GlobCover\\ 
  Avg NDVI 5 years & Average of the NDVI for the 5 years of pre-wildfire periods (averaged over pixels). & LANDSAT\\ 
  Std NDVI 5 years & Standard deviation of the NDVI for the 5 years of pre-wildfire periods (averaged over pixels). & LANDSAT\\ 
  Burning Index & Burning index, a proxy for fire weather hazard, as defined in the NFDRS System (averaged over pixels). & GridMET\\ 
  Maximum Temperature & Maximum Temperature in Kelvin degrees  (averaged over pixels). & GridMET\\
  Rain & Daily precipitation in mm total  (averaged over pixels). & GridMET\\ 
  Solar Radiation & Solar Radiation in W/m$^2$ (averaged over pixels). & GridMET\\ 
  \hline
  \end{tabular}
\end{table}
\renewcommand{\baselinestretch}{1.5} 

\subsection{Satellite Data}
The NDVI is one of the Landsat Surface Reflectance Derived Specral Indices (LSR-DSI). For each pixel in a satellite image, it is defined as
\[
\mbox{NDVI}=\frac{\mbox{NIR} - \mbox{Red}}{\mbox{NIR} + \mbox{Red}},
\]
where Red is the spectral reflectance measurement in the red band of the spectrum (centred near \SI{0.66}{\micro\metre}), and NIR measures the reflectance in the near-infrared band (centred near \SI{0.87}{\micro\metre}). Both, Red and NIR, are codified as 256 grey levels. 
Therefore the values of NDVI are always between $-1$ and $1$, but in general they are non-negative.
Large values of NDVI are associated with high contents of live green vegetation.

For instance, 
Figure \ref{fig:Sample_wildfire_2000-2016.} shows the NDVI (in red, averaged over pixels) for the MEU LIGHTNING COMPLEX (MIDDLE) wildfire example. This area was covered mainly by evergreen forest, having large NDVI values before the wildfire (they oscillate around $0.75$).

We use the GEE platform to obtain the NDVI for images provided by three Landsat satellites (LT5, LT7 and LO8) masking clouds, shadows and snow pixels and removing pixels from water bodies such as lakes, reservoirs, rivers and creeks, as we already did in a previous study \citep{serra2020estimating}. The Landsat satellites provide a consistent source of 30m per pixel resolution, with a frequency of 16 days (approximately 26 observations per year).  All the pixels within a burned region are aggregated by taking the average of each spectral index. In this way a time series of NDVI values is obtained for each region of interest. Further details can be found in \citep{serra2020estimating}.

Given that our main goal is predicting wildfires effects using pre-wildfire observable covariates, two additional explanatory variables were created from the spectral indices data. The average and standard deviation of the NDVI for 5 years of pre-wildfire periods were computed for all observations. 
These two variables work as proxies for the type of vegetation, e.g. larger NDVI values usually show forested areas, whereas lower values of the average of NDVI and larger standard deviations (associated with strong cyclical patterns) indicate grasslands or shrublands types of vegetation. 

In addition, climatological covariates or weather conditions were obtained using GEE from GridMET \citep{Abatzoglou2013}. These were also aggregated on the regions of interest, taking averages over the regions of interest on all the pre-wildfire available periods (from 1990 until the period where each wildfire occurs).  This dataset has a resolution of 4km per pixel and contains the maximum and minimum temperature (in Kelvin degrees), precipitation accumulation (in daily milimetres), downward surface shortwave radiation (in $W/m^2$), and burning index from the National Fire Danger Rating System (NFDRS, \citep{Wildfire2002}). 

\subsection{Effects of Wildfires Data}
The main contribution of \cite{serra2020estimating} was to estimate the effect of the studied wildfires over time.%, for a period of 7 years after the wildfire date. 
The wildfire effect was estimated as the difference between the observed spectral index and the estimated counterfactual (the values that the spectral index would have taken in a hypothetical scenario with the absence of wildfire).
Counterfactuals are estimated in \cite{serra2020estimating} following the  proposals in \cite{athey2021matrix}, a way to perform GSC \citep{Xu2017} based on matrix completion.

\begin{figure}
    %\centering
    %\hspace*{-2cm}
    \includegraphics[width=.9\textwidth]{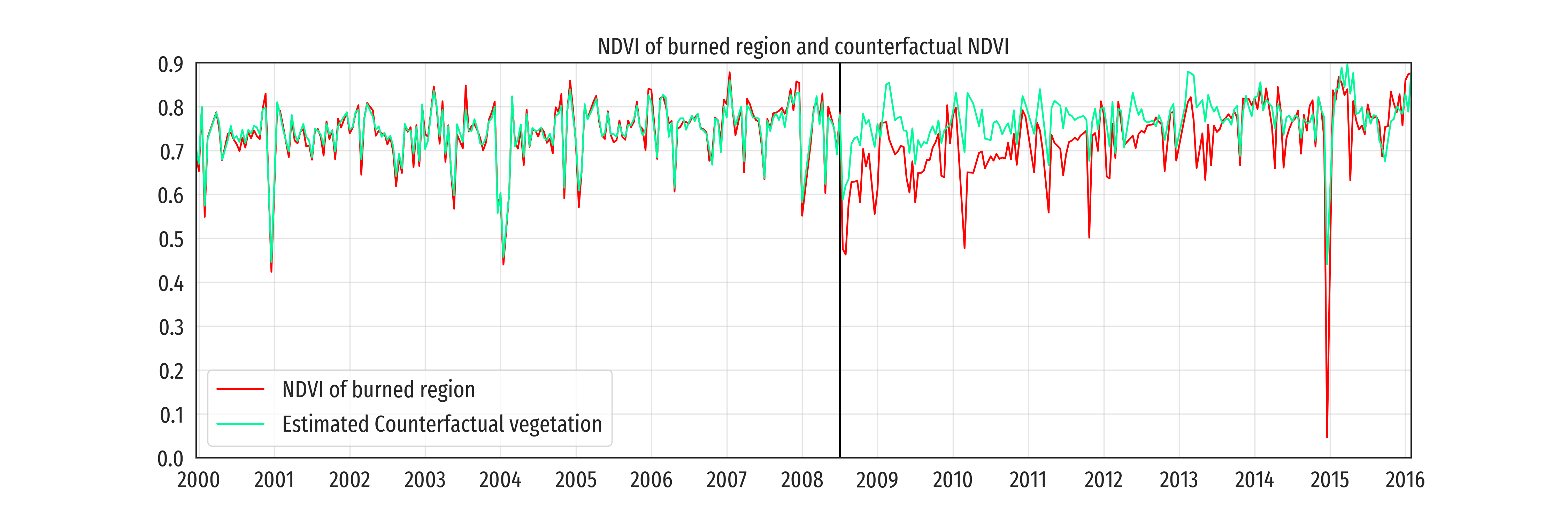}
    \caption{{NDVI for MEU LIGHTNING COMPLEX (MIDDLE) wildfire.} Plot of the vegetation indices NDVI of burned region and counterfactual NDVI vegetation between 2000 and 2016. This figure shows the evolution of NDVI, as well as the counterfactual estimated as explained in \citep{serra2020estimating}.}
    \label{fig:Sample_wildfire_2000-2016.}
\end{figure}

\begin{figure}
    %\centering
    %\hspace*{-2cm}
    \includegraphics[width=.9\textwidth]{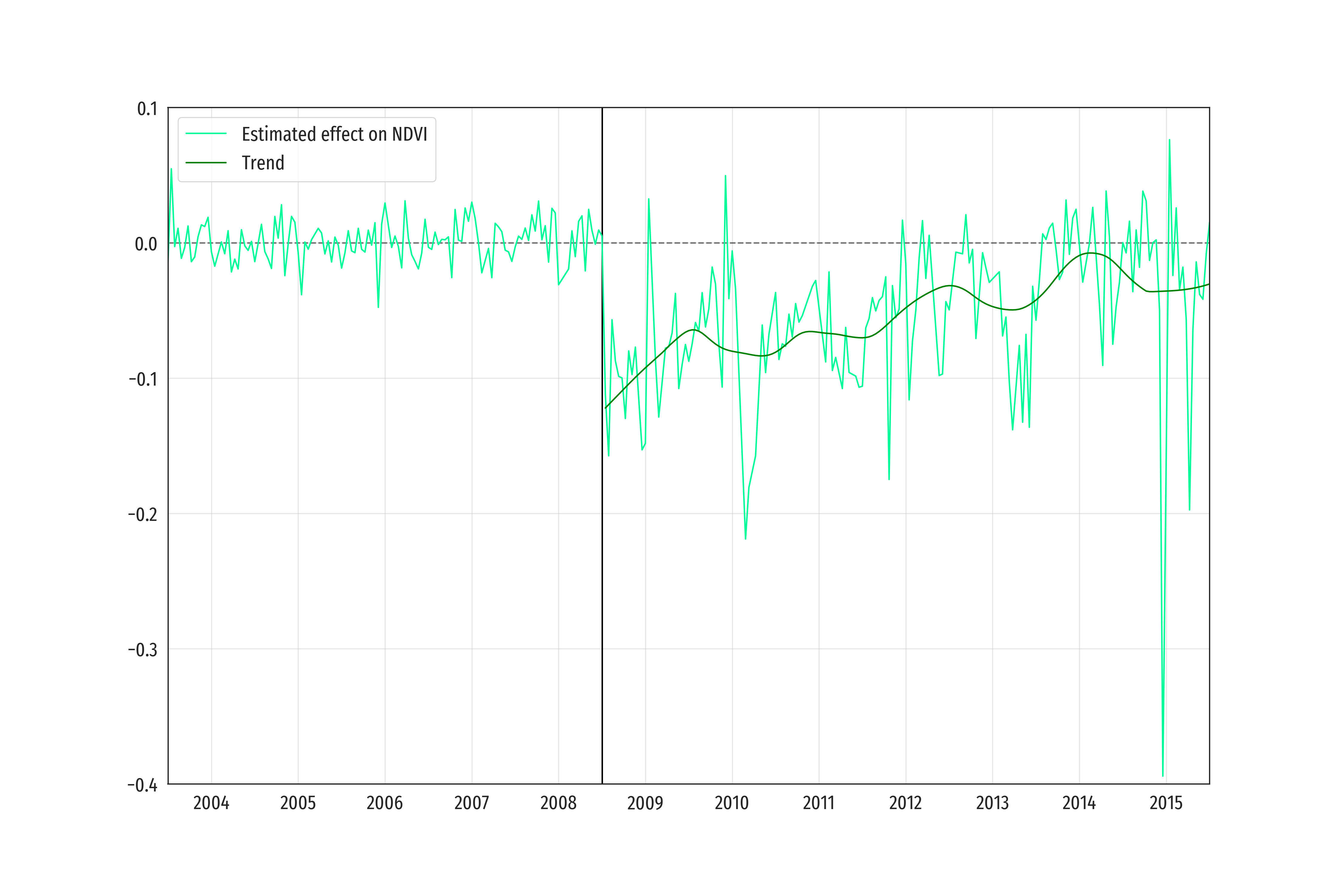}
    \caption{{Plot of the effect and trend extracted for the MEU LIGHTNING COMPLEX (MIDDLE) wildfire.} The estimated effect is shown in light green. The extracted trend is shown in dark green. This figure shows 5 previous years and 7 years after the wildfire, as this is the inclusion criteria in this study.}
    \label{fig:Estimated_effect_and_trend}
\end{figure}

Figures \ref{fig:Sample_wildfire_2000-2016.} and \ref{fig:Estimated_effect_and_trend} illustrate, for the MEU LIGHTNING COMPLEX (MIDDLE) wildfire example, the effect estimation process performed in \cite{serra2020estimating}. 
Figure \ref{fig:Sample_wildfire_2000-2016.} shows the observed NDVI as well as the estimated counterfactual vegetation index. The estimated effect is the difference between these two time series and it is shown in Figure \ref{fig:Estimated_effect_and_trend}.

A descriptive analysis of the estimated wildfires effects is performed in \cite{serra2020estimating}. Among its findings are the following.
Depending on the region burned and the vegetation of these places, the effects can last from less than 2-3 years to more than a decade post-wildfire, and sometimes change the state of vegetation permanently. 
\cite{serra2020estimating} also found that the dynamical effects vary across regions, and have an impact on seasonal cycles of vegetation in later years. 
In order to have more conclusive results than the descriptive ones found in \cite{serra2020estimating}, statistical models must be proposed and estimated. 
A promising possibility is considering regression models with functional response (the estimated wildfire effects as functions of the time elapsed after the wildfire) and explanatory variables such as geographical location, burn severity, size of the burned area, and land cover/vegetation type. 
This constitutes the main contribution of the present project.

The wildfire effects over time estimated in \cite{serra2020estimating} for 7 years post-wildfire and for each of the 243 wildfires that meet our inclusion criteria, are the base from which we construct the functional dataset that will be analyzed in this study. We perform one last step to preprocess the data, that is the trend extraction as explained in Section \ref{subsec:Trend_decomp_subsection}.

\section{Methods}
NDVI time series usually present seasonality, as vegetation changes throughout the seasons of the year. This is especially evident for some types of vegetation, such as grasslands or shrublands. Therefore, we expect post-wildfire NDVI time series of both, the observed and the estimated 
counterfactual vegetation indices, to present seasonal components. These seasonal components will have different amplitudes, since the burned region will present distinct seasonal patterns during the recovery. Therefore, the difference between the burned region NDVI and the counterfactual NDVI will presumably present a changing seasonal pattern. 

Note that, when aligning all the timings of the wildfires, the seasonal pattern of each particular wildfire will present a different phase, as the timings throughout the year of wildfires are different: some wildfires occur on summer periods as opposed to the ones that occur during early spring. Hence, before aligning the recoveries for all wildfires, to conform a unique functional dataset with no mismatches in the phases of seasonality we need to extract the seasonal pattern of each wildfire separately.

In addition, several aspects of the remote sensed data can produce measurement error. Even though pixels that captured clouds were not included at the timing of aggregating multispectral data to measure vegetation, other types of noise could have potentially leaked in the data. To reduce the amount of noise and extract recoveries of vegetation from wildfires, it is suitable for this analysis to smooth the data. 

Therefore, for each time series, we perform a LOESS decomposition, that will simultaneously remove the individual seasonality from the time series, as well as remove noise from the remotely sensed data.

\subsection{Trend Extraction with LOESS and Functional Representation of Data}
\label{subsec:Trend_decomp_subsection}
Once the effects for each wildfire are obtained, we decompose the time series into its structural components. %We want to extract the trend from the data, as different wildfires occurred at different times of the year, and thus might show different seasonal patterns that might hinder the underlying recover that we are trying to estimate and explain. 
Trend extraction of univariate data is a wide field of study \citep{Alexandrov2012}, where the classical decomposition model \citep{brockwell2016introduction} is a time series decomposed in additive terms, separating trend, seasonal component and residuals. 
Assuming the time series can be expressed as the addition of separate terms, for a wildfire starting at calendar time $t_0$ (in years) we have
\[
y(t) = T(t) + S(t) + R(t),    
\]
where $y(t)$ is the outcome observed $y$ at time $t=t_0 + j/26$ for $j \in \{1, \dots, N=7\times 26\}$, $T(t)$ is the trend component at time $t$, $S(t)$ is the seasonal component, which is approximately periodic with cycles of length one year (26 instants of time) in our case, 
%which can be formulated by multiple components, 
and $R(t)$ is the residual component of the time series. 

One method commonly used in many fields for time series decomposition is the Seasonal-Trend decomposition procedure using LOESS \citep{cleveland1990stl}, that is based on local polynomial fitting. This procedure presents several advantages, such as the flexibility on the trend and seasonal components extraction or the ability to decompose series with missing values.

\begin{figure}
    \centering
    %\hspace*{-2cm}
    \includegraphics[width=.9\textwidth]{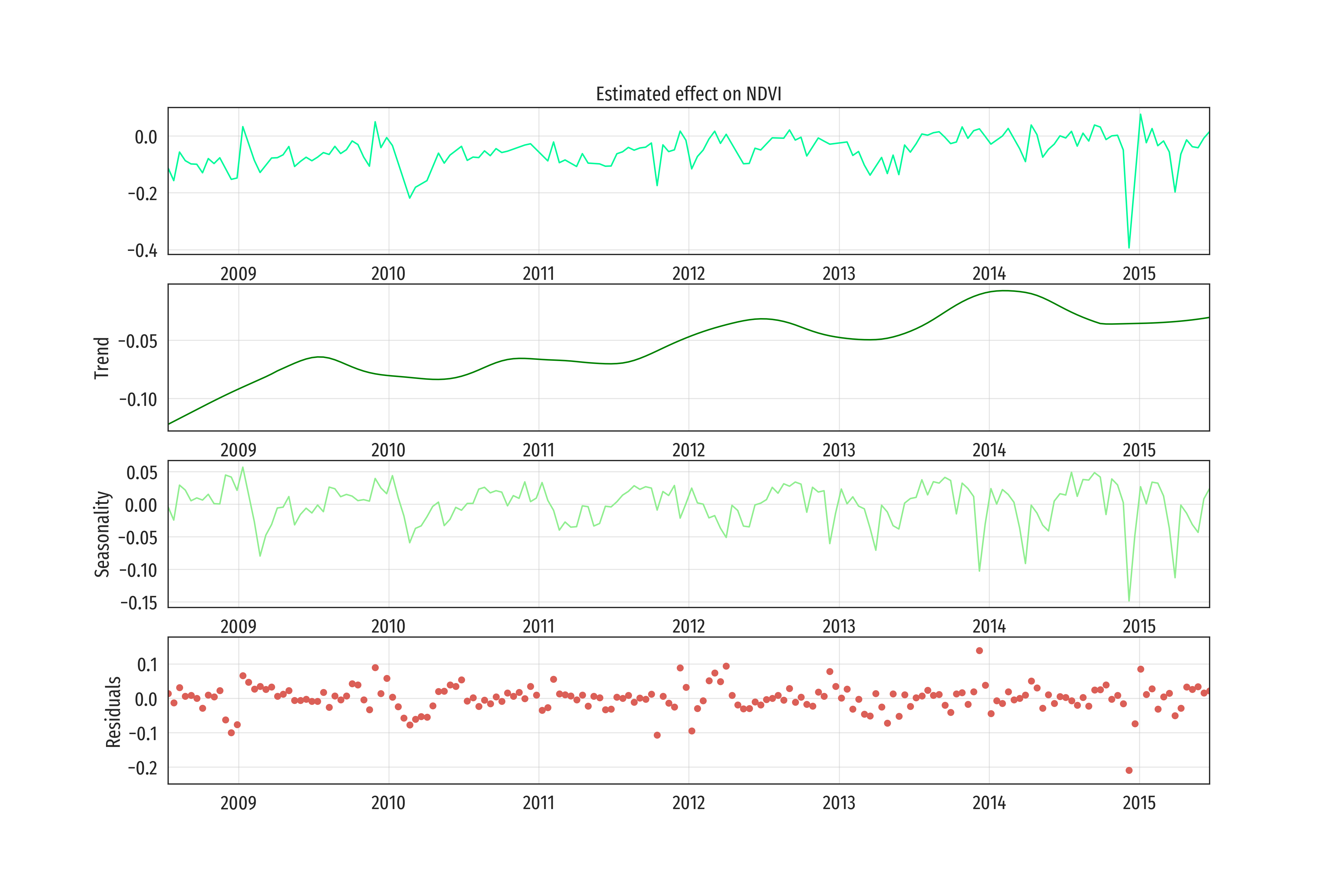}
    \caption{{Plot of time series decomposition using LOESS of the MEU LIGHTNING COMPLEX (MIDDLE) wildfire.} 
    The four graphics show, from top to bottom, the estimated wildfire effect, the extracted trend, the seasonal component, and the residuals.}
    \label{fig:Time_series_decomposition.}
\end{figure}

We use the LOESS implementation from the Python library {\tt statsmodels} \citep{seabold2010statsmodels} to extract the trend from the wildfire effects time series, removing  the seasonal and the residual components at once. 
Figure \ref{fig:Time_series_decomposition.} shows an example of the time series decomposition in the MEU LIGHTNING COMPLEX (MIDDLE) example. Figure \ref{fig:Estimated_effect_and_trend} also shows (in dark green) the extracted trend over the estimated effect (in light green).

\begin{figure}
     \centering
     %\hspace*{-2cm}
     \includegraphics[width=.9\textwidth]{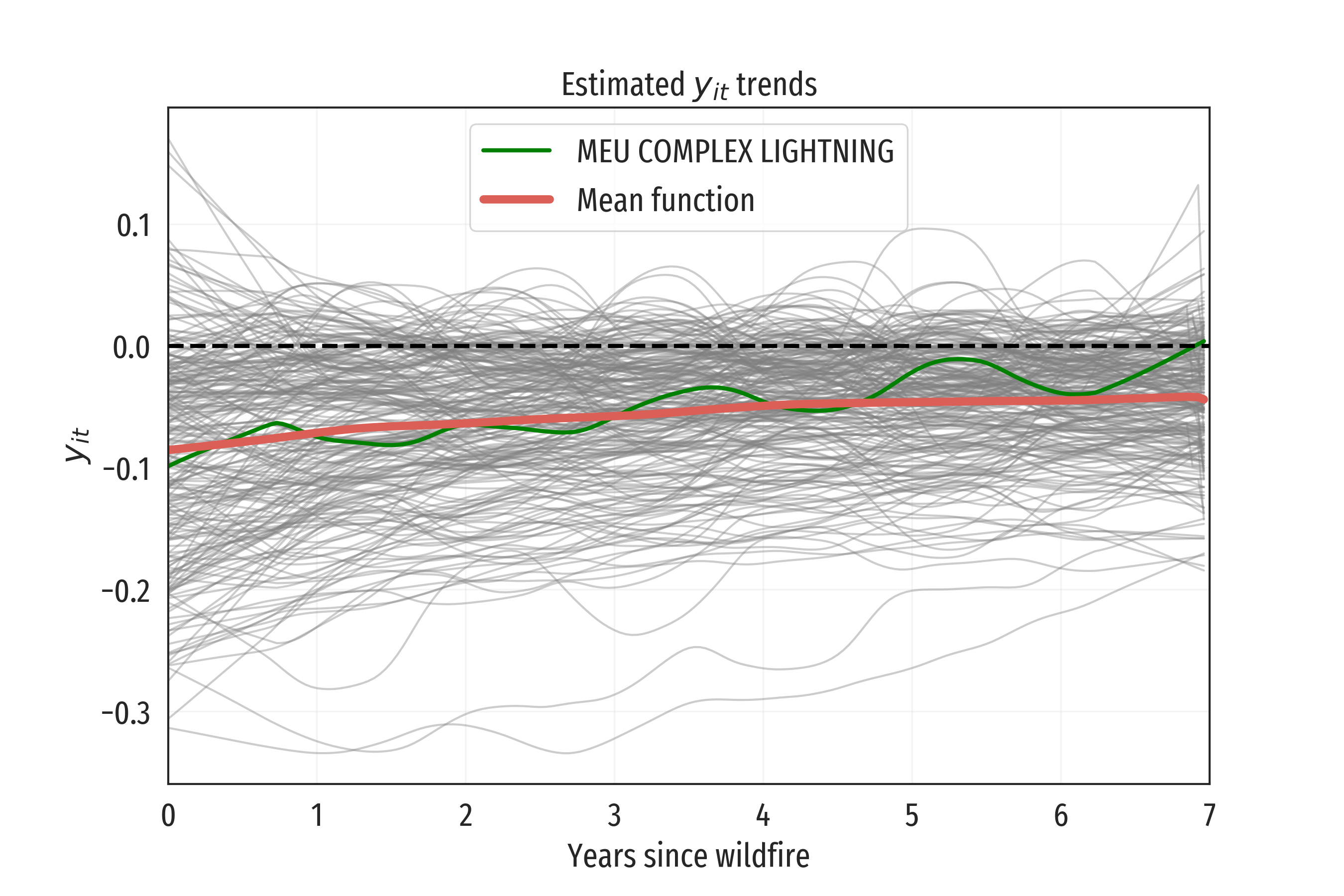}
     \caption{{Plot of the functional dataset,} composed by the extracted trends from the 243 estimated wildfire effects. The trend corresponding to the MEU LIGHTNING COMPLEX (MIDDLE) wildfire example has been marked in green. The functional mean is also represented (in red).}
     \label{fig:Functional_representation_of_data_with_mean}
 \end{figure}

Finally, we align all the extracted trends at $t_0=0$ and represent them as functional data. 
Each of the 243 wildfires is now represented by a function over 7 years of recovery. Each year of data contains 26 discrete values for each observation. Figure \ref{fig:Functional_representation_of_data_with_mean} shows the functional dataset of NDVI trend recoveries, jointly with their mean function. The MEU LIGHTNING COMPLEX (MIDDLE) wildfire example is also highlighted in the figure.

%\begin{figure}
%    \centering
%    \includegraphics[width=.8\textwidth]{Figures/Plot_years_since_wildfire_trends.png}
%  \caption{\textbf{Plot of smoothed data of wildfire recoveries.}}
% \label{fig:Functional_representation_of_data.}
%\end{figure}

\subsection{Functional Principal Components Analysis}
Functional Principal Component Analysis (FPCA; see, for instance, \cite{RamSil:2005} or \cite{HorKok:2012}) is a dimensionality reduction technique for functional data that generalizes the well known Principal Component Analysis extensively used for multivariate data.

Given a functional dataset, FPCA determines the main modes of variation of the observed functions around the mean function.
Formally, FPCA can be stated as follows.
Given a functional dataset $\{y_i(t): i=1,\ldots, n,\, t\in \mathcal{T}=[a,b]\subset \mathbb{R}\}$ with mean function $\bar{y}(t)=(1/n)\sum_{i=1}^n y_i(t)$, 
we look for functions $g_1,\ldots,g_q$ 
({\em principal functions}) 
and real numbers ({\em scores}) $\psi_{ij}$, $i=1,\ldots,n$, $j=1,\ldots,q$, 
such that 
$$
\sum_{i=1}^n \int_\mathcal{T} \left( (y_i(t) - \bar{y}(t)) - \sum_{j=1}^q \psi_{ij}g_j(t)  \right)^2 dt
$$
is minimum. Moreover, the functions $g_1,\ldots,g_q$ are required to be orthonormal \linebreak ($\int_\mathcal{T} g_i(t)g_j(t)dt= \mathbbm{1}_{\{i=j\}}$).
In other words, we are looking for a representation of functional data in 
the $q$-dimensional space spanned by the functions $g_1(\cdot),\ldots,g_q(\cdot)$:
\[%\begin{equation}\label{eq:rho_FPCA}
y_i(t) \approx \bar{y}(t) + \sum_{j=1}^q \psi_{ij}g_j(t), \; t\in \mathcal{T},\; i=1\ldots n.
\]%\end{equation}

It can be proven that the principal functions are the {\em eigen-functions} corresponding to the largest $q$ eigenvalues 
of the {\em sampling covariance operator}, that is,
\[%\begin{equation}\label{eq:eigen}
\int_\mathcal{T} \hat{c}(s,t) g_j(s) ds = \lambda_j g_j(t), \mbox{for all } t\in \mathcal{T}, \, j=1,\ldots,q,\, 
\]%\end{equation}
with $\lambda_1\ge\cdots \ge \lambda_q$.
Moreover the {\em score} of the $i$-th functional data on the $j$-th principal function is $\psi_{ij}=\int_\mathcal{T} (y_i(t)- \bar{y}(t)) g_j(t)dt$.

\subsection{Functional Regression Models}
Analogous to classical regression models, Functional Regression Models (FRM) regress outcomes based on covariates when using functions as either the outcomes or regressors. Hence, FRM take advantage of the nature of time changing variables, either parametrically or non-parametrically. To do so, it can use the functional representation of both regressors and/or outcomes. 

\subsubsection{Function-on-Scalar Regression}\label{sec:FoSR}
% Functional response on scalar regressors
In this research we use the function-on-scalar regression methodology (see, e.g., \citep{RamSil:2005}, \cite{kokoszka2017introduction}, or \cite{goldsmith2015generalized}) as it allows us to understand the relation between the observed outcome over time, with respect to the fixed covariates observed. 
Let $(X, Y)$ be a pair of random variables, where $Y$ is functional and $X = (X_1, \ldots, X_k)$ is a random vector of dimension $k$. 
The linear function-on-scalar regression model for $Y$ given $X=(x_{i1},\ldots,x_{ik})$ is stated as
\begin{equation}\label{eq:fosr_linear}
    Y_i(t) = \beta_0(t) +
    \beta_1(t) x_{i1}
    + \dots + \beta_k(t) x_{ik} \hspace{2pt} + \varepsilon_i(t),
\end{equation}
where $Y_i(t)$ is the functional response over time $t\in \mathcal{T}$ for the observation $i$, $x_{ij}$ is the value of variable $X_j$ in the observation $i$, 
$\beta_0(t)$ is the functional intercept (it is equal to the mean function $\mathbb(Y(t))$ when the $k$ covariates are centered),
$\beta_j(t)$ is the  functional coefficient for the $j$-th covariate $X_j$ for $j\ge 1$, 
and $\varepsilon_i(t)$ is the functional error for the $i$-th observation, a zero mean continuous stochastic process, assumed to be indpendent for different observations.
The problem of variable selection in the linear function-on-scalar regression model was addressed in \cite{chen2016variable}.

However, different kinds of covariates can be considered, as not all of them have a changing effect over time, or might have different effects. In order to allow the function-on-scalar regression model to admit richer covariate terms, \cite{scheipl2015functional} introduced the 
functional additive mixed model (where functional covariates are also allowed).
As an example, the following equation shows a function-on-scalar additive regression model with terms of different types:
\begin{equation}     \label{eq:fosr_additive}
    y_i(t) = \beta_0(t) + \beta_1x_{i1} + s_2(x_{i2}) + 
    \beta_3(t)x_{i3} + \gamma_4(t,x_{i4}) + \varepsilon_i(t),
\end{equation}
where 
$\beta_0(t)$ is the functional intercept, 
$\beta_1$ is constant over time, %as well as $x_{i1}$, 
$s_2(x_{i2})$ is a smooth function of the covariate, 
$\beta_3(t)x_{i3}$ is the same kind of covariate-coefficient relation from equation (\ref{eq:fosr_linear}), 
$\gamma_4(t, x_{i4})$ is a smooth function depending on $t$ and $x_{i4}$, 
and finally $\varepsilon_i(t)$ is the $i$-th error function. 
Variable selection is less developed for the function-on-scalar additive model than for the linear function-on-scalar model.

\section{Results}
In this section we present the main results from this study, showing how the characteristics of the vegetation and land cover previous to the wildfire, as well as the prior weather conditions to the wildfire, affect the vegetation recovery patterns. 

We start summarizing the functional dataset containing the 243 wildfire recoveries. Their mean function is represented in Figure \ref{fig:Functional_representation_of_data_with_mean}, jointly with the complete dataset. 
The mean wildfire effect on NDVI is always negative for the 7 year period after the wildfire, and the absolute value of this negative effect is monotonically decreasing over time, 
going from $-0.0856$ at time 0 to  $-0.0418$ seven years later, in terms of lost NDVI points, with a global average of $-0.0567$.
In average, the burned areas are progressively recovering $0.0438$ NDVI points after wildfires (approximately 10\% of the range of the functional data set values, see Figure \ref{fig:Functional_representation_of_data_with_mean}). 
It is also noticeable that, on average, it takes more than 7 years for a complete recovery of the NDVI: the value of the mean function after 7 years is still negative.
The library {\tt fda.usc} \citep{Febrero-Bande2012} in R \cite{R.2020} has been used for the descriptive analysis, including the choice of the MEU LIGHTNING COMPLEX (MIDDLE) as an illustrative wildfire example, as it has the modal median recovery function in 2008 (the modal year).

\begin{figure}
    \centering
    \includegraphics[width=.6\textwidth]{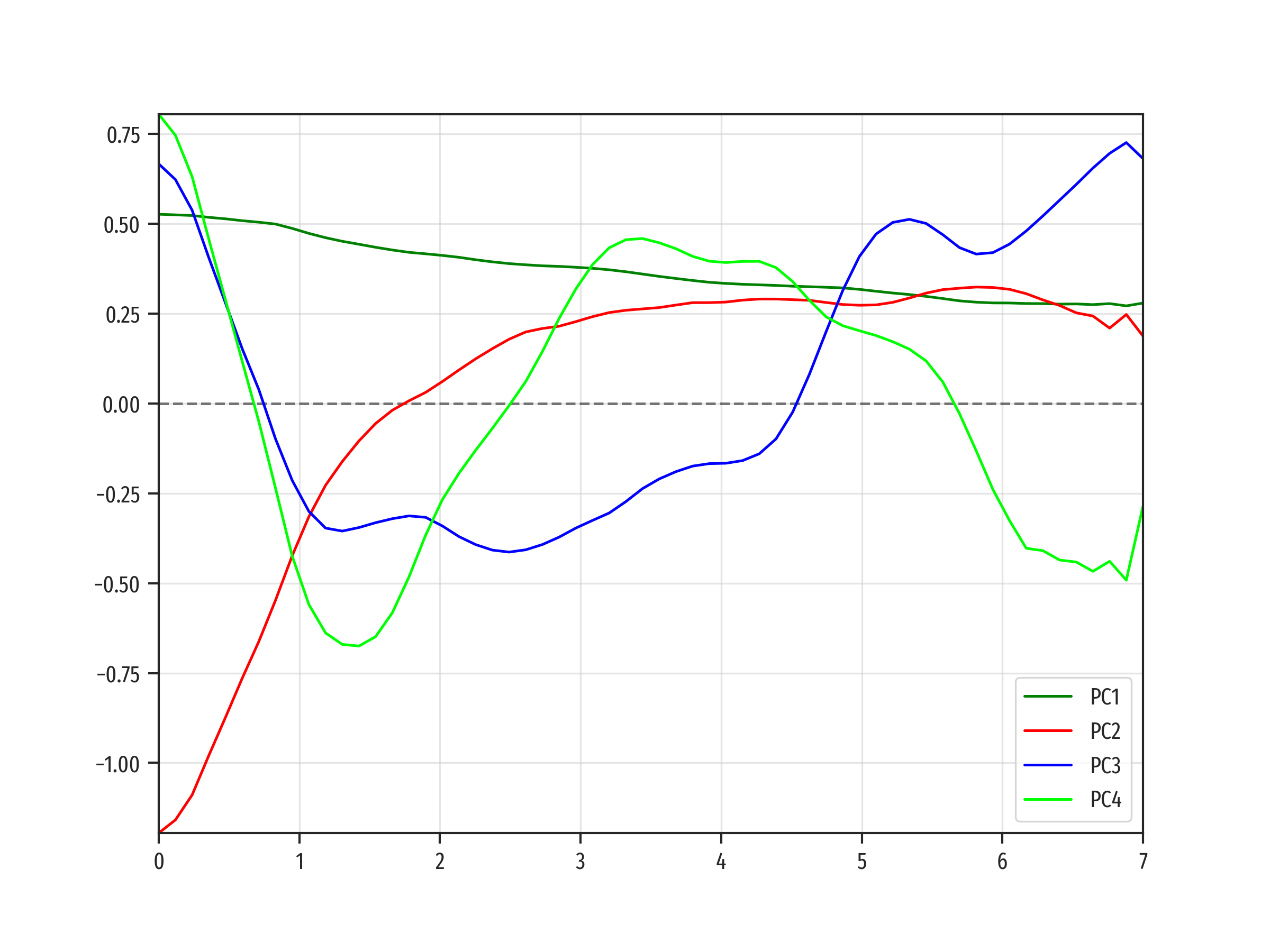}
    \\
    \includegraphics[width=.7\textwidth]{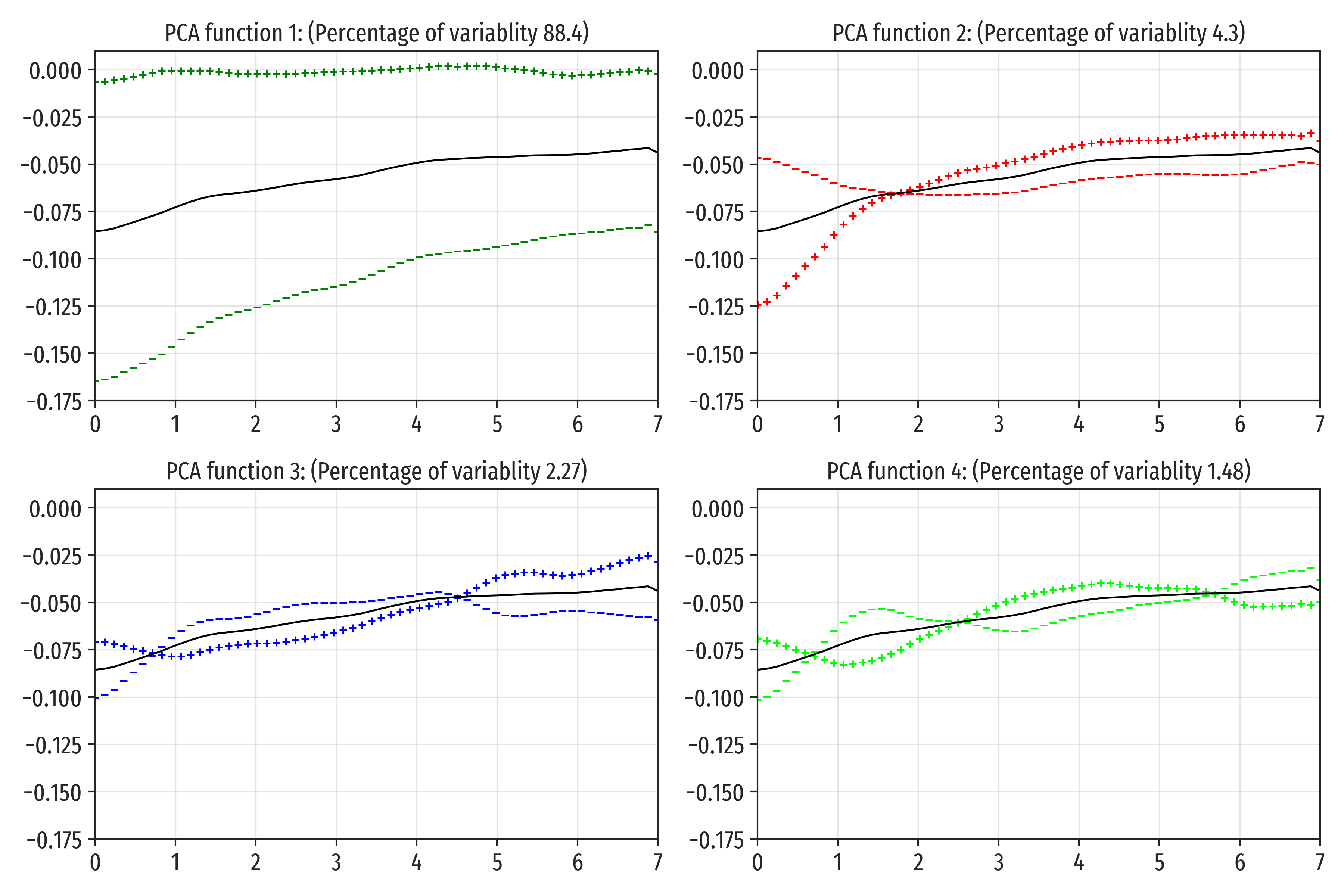}
    \caption{{Functional principal component analysis results for the wildfire recoveries dataset.} 
    The upper plot shows the first four principal functions. The lower plots show the mean (black solid line), and the mean plus/minus a constant times each principal function.}
    \label{fig:Functionalprincipalcomponents}
\end{figure}

Next, FPCA has been applied to find the main modes of variation of the studied functional data around the average. Fig~\ref{fig:Functionalprincipalcomponents} shows the mean, and the mean plus/minus a constant times the first four principal functions, that have been computed 
%using \citep{Febrero-Bande2012}. 
using the function {\tt pca.fd} from package {\tt fda} \cite{fda.package} in R.

The first principal function explains almost $90\%$ of the variability, showing a direction of severity in the NDVI drop: wildfires with positive scores in this principal function experiment smaller drops in NDVI than those having negative scores.
The second principal function ($4.3\%$ of the total variability) can be interpreted as a direction separating wildfires with faster recoveries (those with more positive scores) from those with slower regeneration capacity (wildfires with more negative scores). The following two functional components only explain less than $4\%$ of the total variance, with no clear recovery patterns, so they should be interpreted with caution. 

The main goal of this study is to quantify the influence that different pre-wildfire conditions (geographical region, climatological conditions, or vegetation types) of the burned areas have on wildfire effects over the subsequent years post-wildfire. 
In order to achieve this goal, function-on-scalar additive models (of the type from equation \ref{eq:fosr_additive}) are fitted using the function {\tt pffr} from the library {\tt refund} \cite{refund.package} in R. 
The list of potential covariates to be included in this model is given in Table \ref{tab:list_of_covariates}.

As far as we know, the variable selection problem for the function-on-scalar additive model is still an open issue, as we mentioned in Section \ref{sec:FoSR}. 
In fact, library {\tt refund} includes a function doing variable selection for the linear function-on-scalar model ({\tt fosr.vs}), but not for the additive extension. 
Additionally, each of the explanatory variables can enter in the function-on-scalar additive model in several ways, as it is illustrated in equation (\ref{eq:fosr_additive}). 
Therefore we have developed a heuristic model building strategy, which we describe below.

To select the way in which we introduce each covariate to the function-on-scalar additive model, five different univariate models have been fitted for each covariate separately.
Exceptions were made for three pairs of covariates (longitude and latitude, average and standard deviation of NDVI during 5 years pre-wildfire, and landcover and landcover entropy) that have been included together additively in these 5 single models, because both variables in each pair are jointly summarizing the same characteristic (geographic location, NDVI, and land cover).
Table \ref{tab:devexpl} shows the results from the $11\!\!\times\!\!5=55$ different fitted models (all of them being sub-models of equation (\ref{eq:fosr_additive})), in terms of the percentage of observed variability explained (100 times the adjusted $R^2$).

\renewcommand{\baselinestretch}{.8} 
\renewcommand{\arraystretch}{1.5}
\begin{table}%[h]
    \caption{Percentage of observed variability explained from $11\!\!\times\!\!5$ univariate or bivariate function-on-scalar regression models.} 
    \label{tab:devexpl}
    \centering
    \begin{tabular}{r|ccccc}
    \multicolumn{6}{r}{\bf Term included in each model \hspace*{1cm} }\\
    \textbf{Variable} & $\beta x$ & $s(x)$ & $\beta(t) x$ & $\beta(t) x + s(x)$ & $\gamma(t,x)$ \\
    \hline
  Latitude, Longitude & 7.05 & \textbf{19.91} & 7.08 & 19.94 & 17.63 \\ 
  Avg Elevation & 19.03 & \textbf{23.86} & 19.32 & 24.15 & 24.91 \\ 
  Year & 4.44 & \textbf{7.93} & 4.50 & 7.99 & 7.19 \\ 
  Start Month & 6.40 & \textbf{7.72} & 6.57 & 8.39 & 8.17 \\ 
  log(Acres) & 6.20 & \textbf{8.91} & 6.51 & 9.22 & 9.02 \\ 
  \begin{tabular}{r} Landcover and\\[-.3em] Landcover Entropy\end{tabular}\hspace*{-.2cm}
  & 4.92 & \textbf{7.25} & 4.72 & 7.26 & 6.98 \\ 
  \begin{tabular}{r}Avg and Std \\[-.3em] NDVI 5 years before\end{tabular}\hspace*{-.2cm} & 30.58 & 43.98 & 33.45 & \textbf{46.85} & 46.98 \\ 
  Burning Index & 8.71 & \textbf{14.70} & 9.00 & 14.99 & 13.52 \\ 
  Maximum Temperature & 21.74 & \textbf{27.78} & 22.19 & 28.23 & 28.93 \\
  Rain & 22.17 & 29.22 & 23.25 & \textbf{30.30} & 28.34 \\ 
  Solar Radiation & 7.60 & \textbf{15.18} & 7.70 & 15.28 & 16.24 \\ 
  \hline
    \end{tabular}
\renewcommand{\baselinestretch}{1.5} 
\end{table}

The columns in Table \ref{tab:devexpl} correspond to different types of models, and the rows to the variable (or to the pair of variables) used as regressors in the models. 
In each row, the complexity of the models increases from left to right:
in the first two models, the terms depend only on the explanatory variable (linearly first, then non-parametrically), while in the other three models it depends on both, the covariate and the time index (in the third column, the term is linear in the covariate and nonparametric in time, 
the fourth model includes the second and third models terms additively, and finally the fifth model is nonparametric simultaneously in  the covariate and the time index).
In general, the models including a nonparametric term in the covariates have larger percentages of explained variability (columns 2, 4 and 5, which show an even performance) than those that are linear in the covariates (columns 1 and 3). 
Additionally, the inclusion of time dependent coefficients $\beta(t)$ (column 3) does not represent a large improvement with respect to the standard linear term (column 1).
Therefore, for each row, a model has been selected according to a balance between explanatory power and model simplicity: a simpler model is preferred to a more complex one, if the difference in percentage of explained variability is less than 1\%.
At each row, the selected model is marked in bold.

Observe that the best univariate (or bivariate) fits in Table \ref{tab:devexpl} correspond to the models having average and standard deviation of NDVI for the 5 previous years to the wildfires as covariates (almost $47\%$ of explained variability), followed by those including rain ($30\%$) or maximum temperature (around $28\%$) as explanatory variables. 

Despite we do not delve any further into the results of these simple models (further comments on individual covariates effect on the response will be made below), we are going to build a multiple function-on-scalar additive model.
Rather than delving further into the results of these simple models, we are going to build an additive multiple function scalar model, which in turn will provide further insights on the effect of individual covariates on the response.

\begin{figure}%[h]
    \centering
    \includegraphics[width=.7\textwidth]{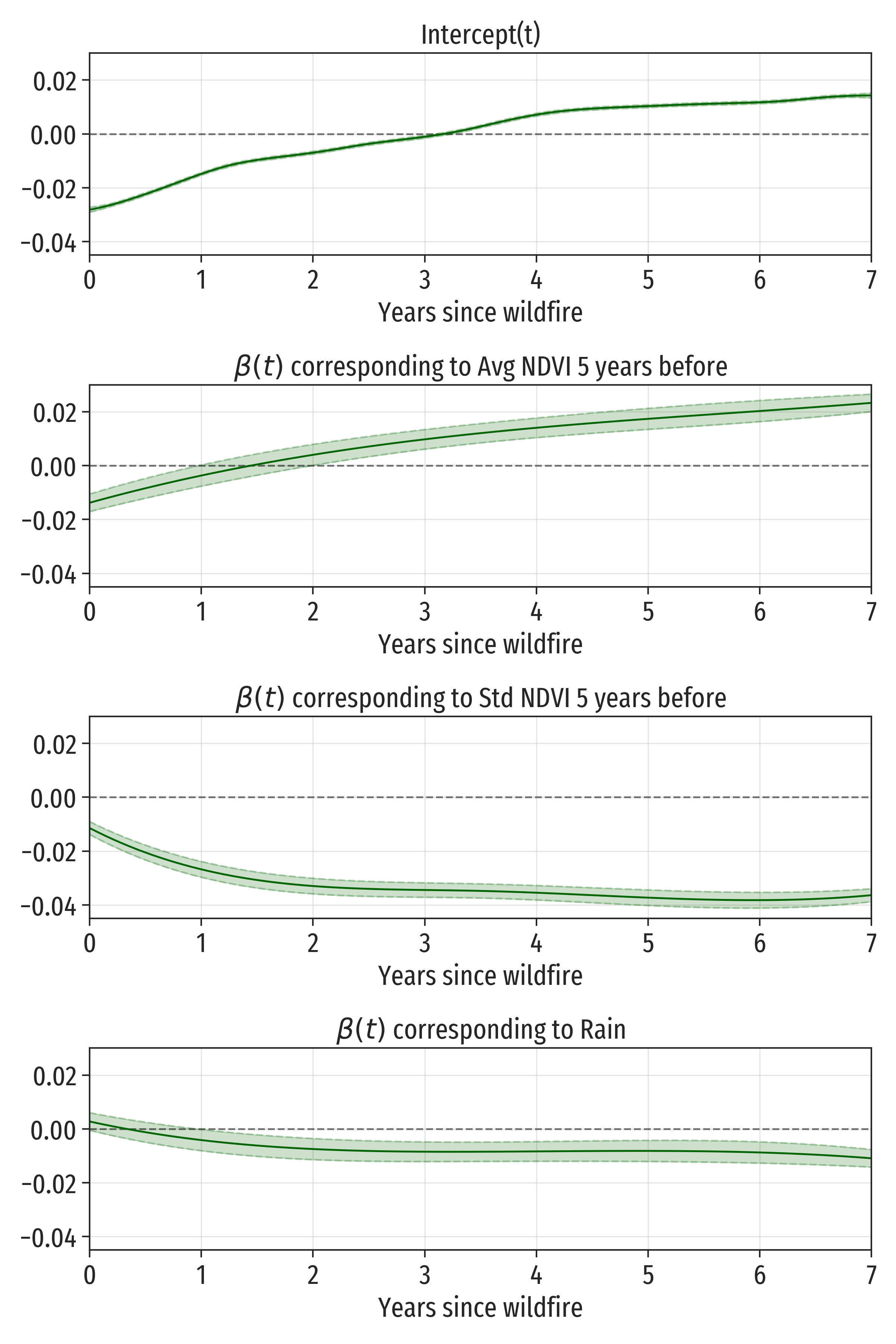}
    \caption{Full function-on-scalar additive model. Estimated functional coefficients of the form $\beta_j(t)$.}
    \label{fig:fosr_cov_over_time}
\end{figure}

\begin{figure}%[h]
    \centering
    \includegraphics[width=.7\textwidth]{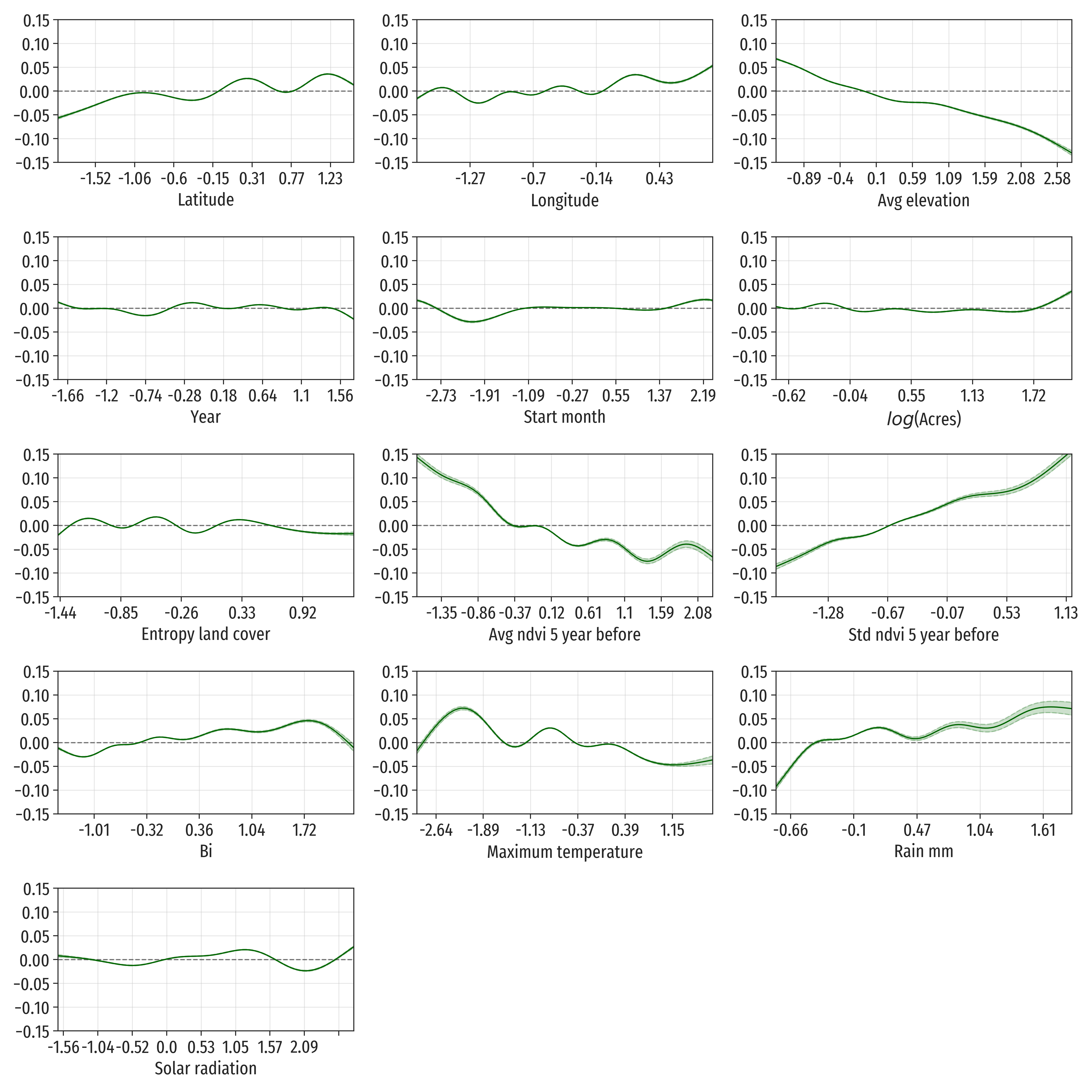}
    \caption{Full function-on-scalar additive model. Estimated smooth terms of the form $s_j(x_j)$.}
    \label{fig:fosr_cov}
\end{figure}

We then proceed to fit a full model (using again the function {\tt pffr} in {\tt refund}), which includes the terms selected in Table \ref{tab:devexpl}.
The covariates have been centered and standardized before fitting the model to force all of them to share a common scale. 
This way the estimated functions are comparable to each other. 
Tables \ref{tab:FoSR.full.2} and \ref{tab:FoSR.full.1}, and 
Figures \ref{fig:fosr_cov_over_time} and \ref{fig:fosr_cov},
summarize the fitted model.
This model explains a $72.9\%$ of the variability observed in the response, strongly improving the best model included in Table \ref{tab:devexpl} ($46.98\%$).
Tables \ref{tab:FoSR.full.2} and \ref{tab:FoSR.full.1} indicate that all the terms included in the model are highly significant. This fact and the large percentage of explained variability suggest that this is an adequate model. 

\begin{table}
\renewcommand{\baselinestretch}{.8} 
\renewcommand{\arraystretch}{1.5}
\centering
\caption{Full function-on-scalar additive model. Estimation of the parametric terms.}
\label{tab:FoSR.full.2}
\centering
\begin{tabular}{lcccc}
\textbf{Parametric terms}           & \textbf{Estimate}      & \textbf{Std. Error} & \textbf{t value}        & \textbf{Pr($\geq |t|$)}   \\
\hline
(Intercept)                                     & -0.0574    & 0.0003548  & -155.253   & $<$ 2e-16     \\
Landcover Grassland/Herbaceous          & -0.0022    & 0.0003699  & -4.085     & 4.42e-05  \\
Landcover Shrub/Scrub                   &  0.0031    & 0.0004906  & 6.338      & 2.34e-10  \\
Landcover Other               &  0.0046    & 0.0013508  & 3.429      & 0.000606  \\
\hline 
\end{tabular}

\ \\ \  \\ \ 

\caption{Full function-on-scalar additive model. Estimation of the  nonparametric terms.}
\label{tab:FoSR.full.1}
\begin{tabular}{lcccc}
\textbf{Nonparametric terms}           & \textbf{Edf}      & \textbf{Ref.df} & \textbf{F}        & \textbf{p-value}   \\
\hline
Intercept(t)                                    & 13.218    & 19.000    & 364.17      & $<$ 2e-16     \\
$s_1$(Latitude)                                  & 8.973     & 9.000     & 346.12      & $<$ 2e-16     \\
$s_2$(Longitude)                                 & 8.984     & 9.000     & 321.41      & $<$ 2e-16     \\
$s_3$(Avg Elevation)                             & 8.632     & 8.960     & 520.30      & $<$ 2e-16     \\
$s_4$(Year)                                      & 8.959     & 8.999     & 123.46      & $<$ 2e-16     \\
$s_5$(Start Month)                               & 4.977     & 5.000     & 85.87       & $<$ 2e-16     \\
$s_6$(log(Acres))                                & 8.906     & 8.997     & 9.000       & $<$ 2e-16     \\
$s_7$(Entropy landcover)                         & 8.977     & 9.000     & 346.38      & $<$ 2e-16     \\
$s_{8}$(Avg NDVI 5 years before)                   & 8.988     & 9.000     & 675.16      & $<$ 2e-16     \\
$\beta_9(t)$ Avg NDVI 5 years before            & 3.558     & 3.831     & 377.16      & $<$ 2e-16     \\
$s_{10}$(Std NDVI 5 years before)                   & 8.825     & 8.989     & 102.17      & $<$ 2e-16     \\
$\beta_{11}(t)$ Std NDVI 5 years before           & 3.962     & 3.999     & 433.37      & $<$ 2e-16     \\
$s_{12}$(Burning Index)                             & 8.940     & 8.998     & 214.318     & $<$ 2e-16     \\
$s_{13}$(Maximum temperature)                       & 8.980     & 9.000     & 487.19      & $<$ 2e-16     \\
$s_{14}$(Rain)                                      & 8.966     & 8.999     & 286.88      & $<$ 2e-16     \\
$\beta_{15}(t)$ Rain                           & 3.585     & 3.844     & 32.83       & $<$ 2e-16     \\
$s_{16}$(Radiation)                                 & 8.923     & 8.998     & 249.83        & $<$ 2e-16     \\
\hline 
\end{tabular}
\renewcommand{\baselinestretch}{1.5} 
\end{table}
%. R-sq.(adj) =  0.729;   Deviance explained = 72.9\%; -REML score = -90628;  Scale est. = 0.0009497:  n = 44226 (243 x 182)

We describe first the results for the parametric part of the model (Table \ref{tab:FoSR.full.2}), which only includes the covariate Landcover (a factor with 4 levels) with constant effects over time. 
The reference level for this factor is {\em Evergreen forest}. Table \ref{tab:FoSR.full.2} shows that burned areas having had Grassland/Herbacious as dominant land cover experiment larger decrement in NDVI than evergreen forest areas. The opposite happens for areas at which shrubland or scrubland were dominant. 
Regarding the constant coefficients, the most affected areas when a wildfire happens are grassland/herbaceous (that loose 0.0596 points of NDVI in average; we noted before that the global average loss is 0.0567 NDVI points), followed by evergreen forests (loosing 0.0574 points of NDVI), then shrublands and scrublands (with a reduction of 0.0543 points of NDVI), and finally areas at which other types of vegetation are dominant (where the NDVI reduction is of 0.0528 points in average). 
However, the landcover covariate cannot be interpreted separately from the other covariates (mainly the average and the standard deviation of NDVI, which strongly depend on types of landcover). 

We move our attention now to non-parametrically estimated terms, using the information contained in Table \ref{tab:FoSR.full.1} and in Figures \ref{fig:fosr_cov_over_time} (showing the estimation of the functional coefficients $\beta_j(t)$) and \ref{fig:fosr_cov} (which include the estimations of the functions $s_j(x_j)$). 

The estimation of the function $\beta_0(t)$ in model (\ref{eq:fosr_additive}) is labeled {\tt Intercept($t$)} in Figure \ref{fig:fosr_cov_over_time} (upper panel). Except for a vertical shift, it is approximately equal to the mean function (see Figure \ref{fig:Functional_representation_of_data_with_mean}).
The vertical shift should be equal to the estimated {\tt Intercept} in Table \ref{tab:FoSR.full.2} if there were no factor covariates in the model. In our case, however, this {\tt Intercept} is referred to the level {\em Evergreen forest} of the factor Landcover. 

There are three covariates (Avg NDVI 5 years before, Std NDVI 5 years before, and Rain) that contribute with two terms ($\beta_j(t) x_j$ and $s_j(x_j)$) to the full additive function-on-scalar model.
To understand the contribution of these variables to the response recovery functions, we have to consider simultaneously the two corresponding estimated functions, where one is represented in
Figures \ref{fig:fosr_cov_over_time} and the other one in Figure \ref{fig:fosr_cov}.
Regarding Avg NDVI 5 years before (average of NDVI over the 5 years before the wildfire), the estimation of its functional coefficient $\beta_j(t)$ (Figure \ref{fig:fosr_cov_over_time}, second panel) presents a monotonically increasing pattern with a total increment of $0.04$ NDVI points over the 7 years. At the same time, the estimation of its term $s_j(x_j)$ (Figure \ref{fig:fosr_cov}, third row, second column) is a roughly decreasing function with a range of values of more than $0.20$ NDVI points. So it follows that the contribution of the term $s_j(x_j)$ is much larger than that of the term $\beta_j(t) x_j$ for this explanatory variable. 
The nonparametric term $s_j(x_j)$ indicates that larger values of NDVI vegetation tend to suffer more from wildfires. 
For instance, in average, an area with pre-wildfire NDVI value equal to the mean plus one standard deviation loses $0.1$ NDVI points more than another area with pre-wildfire NDVI value one standard deviation below the average. 
For these two fictitious areas, the effect of the term $\beta_j(t) x_j$ is to add or subtract, respectively, the estimated coefficient $\beta_j(t)$. Then the area with NDVI values over the mean will have a larger decrease in NDVI the first one and a half yeas, but its recovery will be faster than in the area with previous lower NDVI values.

For Std NDVI 5 years before (standard deviation of NDVI over the 5 years before the wildfire), the relative relevance of the term $\beta_j(t) x_j$ is also much smaller than that of the term $s_j(x_j)$: their ranges are $0.03$ and $0.25$, respectively. 
The functional coefficient $\beta_j(t)$, negative for all $t$, is decreasing the first two years and almost constant from then on (with an approximate value of $-0.04$ NDVI points). 
The term $s_j(x_j)$ in this case is an increasing function on the standard deviation of pre-wildfire NDVI values, indicating that vegetation diversity (large values of Std NDVI 5 years before) is a protecting factor against wildfire effects. 
Combining both terms, the difference in loss of NDVI points between two areas with values of Std NDVI 5 years before one standard deviation over and below the mean, respectively, for $t$ larger than two years is
\[
\left(\beta_j(t) + s(1) \right) - \left(\beta_j(t) + s(1) \right) 
\approx (-0.04 + 0.10) - (0.04 - 0.03)= 0.05.
\]
For $t$ smaller than 2 years, the differences between these two areas are smaller than $0.05$ and increasing in $t$.

For the explanatory variable Rain, the term $\beta_j(t) x_j$ is even less important than in the two previous cases (the range of $\beta_j(t)$ is smaller than $0.02$ NDVI points, and it is almost constant from two years after the wildfire). 
On the other hand, the term $s_j(x_j)$, that has an approximate range of $0.17$, grows rapidly at low values of the variable Rain (smaller than $0.3$ times the standard deviation below the mean, approximately) and then it is almost constant or slightly increasing.
We conclude that moderate or large precipitations seem to help recover or protect against the wildfire effects.  

The remaining 10 explanatory variables contribute to the full additive function-on-scalar model only with a nonparametric term $s_j(x_j)$ that remains constant over time after the wildfire. 
The estimations of these terms are represented in Figure \ref{fig:fosr_cov}. 
The most relevant contribution to the model is that of the covariate Avg Elevation, which estimated term $s_j(x_j)$ has a range of $0.20$ NDVI points. This function is decreasing in elevation, indicating that the wildfire effects are larger in more elevated areas, probably because elevated areas present in average richer vegetation (larger pre-wildfire NDVI values) than those with lower elevation. 

Less important, although also worth mentioning, are the explanatory variables Bi (burning index) and Maximum temperature. 
For the burning index, the estimated term $s_j(x_j)$ is an slightly increasing function in the middle part of the range of burning index values. It follows that areas with lower fire hazard will have slightly larger wildfire effects. 
The estimated term $s_j(x_j)$ for maximum temperature is roughly decreasing in its argument, indicating that low maximum temperatures protect moderately against the wildfire effects.

Regarding geographical coordinates contribution to the model, the wildfire effects in the South (respectively, West) are larger than in the North (respectively, East), but the differences are small (less than $0.1$ NDVI points). 

Finally, we do not find clear and strong interpretable patterns of dependence between the response, the wildfire effects functions, and the rest of covariates (Year, Start month, log(Acres), Entropy land cover, and Solar radiation).
%One thing to notice is that we find that wildfires starting at early spring seem to have a larger effect than those that occur in other months. 
%Nonetheless, we leave the results to be interpreted by ecological experts.

\section{Conclusions and Discussion}

The functional regression methodology has shown to be an effective way to study and explain vegetation recovery from wildfires, using pre-wildfire explanatory variables. 
The additive function-on-scalar fitted model explains $72.9\%$ of the total variability of the responses. A large part of the explanatory power of the model goes directly to explain the recovery dynamic through the presence of regression coefficients that change over time. 
Nevertheless, the main part of the relationship between the explanatory variables and the wildfire effects functions is constant over time after the wildfire and, it is worth mentioning, non-linear.

The most important lessons we draw from this model are the following. 
In average, the recovery process after a wildfire is slow and takes more than 7 years (the time span used in this study).
Each particular wildfire is a combination of a unique set of conditions that alter vegetation and ecosystems in a different manner, and it seems that all of them have an effect on the wildfire recovery process.
The main risk conditions for a given area from suffering larger wildfire effects are, in this order, 
to have a rich and homogeneous vegetation (large and uniform NDVI, dominance of grassland, herbaceous vegetation or evergreen forest as land cover),
to present a low precipitation regime, 
to have a large elevation over the sea level, 
to have low burning index,
to have large maximum temperatures, and
to be located in the South or West of California.

The convenience of studying outcomes changing over time, together with the estimation of the effect of several kinds of conditions pre- and post-wildfire, makes functional regression models to be a perfect methodology for this kind of studies. 
Previous studies use standard multiple regression models to compare absolute values of spectral indices, or comparisons of geolocated rasters such that these can include the spatial component of wildfires. However, giving estimates of the effect of these characteristics on the recovery pattern of vegetation from wildfires will allow environmental scientists and land management entities to study the characteristics that need more preservation. 

It is important to notice that this methodology has only been implemented over the recoveries estimated from \citep{serra2020estimating}. Nevertheless, this could be applied in many other research areas and fields, benefiting from the temporal component that this methodology includes, as everything is observed and measured over time. Expanding the study area to other fire-prone regions around the world, and increasing the time-span observed after wildfires (e.g. 15 years after each fire) would probably allow to observe full recoveries from wildfires. However, this remains outside the scope of this work. 

This study tries to close the gap between satellite remote sensing and evaluation of wildfires' effects over time. 
It must be noted that gathering and pre-processing data, usually coming from different sources, is a crucial and highly sophisticated task when dealing with remote sensing data. 
Functional Data Analysis, and functional regression in particular, is an advanced statistical methodology well suited to analyze such rich data sets.

\section{Acknowledgements}

Serra-Burriel would like to thank the Barcelona Supercomputing Center for the Severo Ochoa Mobility Grant, and Delicado would like to thank the Spanish Ministerio de Ciencia e Innovaci\'on for the grant MTM2017-88142-P.

The code used in this work has been performed using Python 3.8.1~\cite{10.5555/1593511} and R 3.6.2~\cite{R.2020} programming languages and the Google Earth Engine (GEE) platform \cite{gorelick2017google}. We would also like to acknowledge the following software libraries used in the analysis:  
fda~\cite{fda.package} (R),
fda.usc~\cite{Febrero-Bande2012} (R),
gsynth~\cite{Rgsynth} (R),
refund~\cite{refund.package} (R),
geopandas~\cite{kelsey_jordahl_2020_3946761} (Python), 
matplotlib~\cite{hunter2007matplotlib} (Python), 
numpy~\cite{harris2020array} (Python), 
pandas~\cite{mckinney2011pandas} (Python), 
scipy~\cite{virtanen2020scipy} (Python).

% ``I always thought something was fundamentally wrong with the universe'' \citep{adams1995hitchhiker}

\bibliographystyle{plainnat}

\bibliography{references.bib}

\end{document}